\documentclass[aps,prl,preprint,showpacs,superscriptaddress,tightenlines]{revtex4}

\usepackage{graphicx}
\usepackage{dcolumn}
\usepackage{bm}
\usepackage{epsfig}

\graphicspath{{ps}}

\begin{document}

\preprint{\vbox{
\hbox{\hfil hep-ex/0503013}
\hbox{\hfil Belle Preprint 2005-7}
\hbox{\hfil KEK Preprint 2004-103}
}}



\title{\quad\\[0.5cm] Measurement of Polarization and 
Triple-Product Correlations in $B \to \phi K^{*}$ Decays}

\affiliation{Budker Institute of Nuclear Physics, Novosibirsk}
\affiliation{Chiba University, Chiba}
\affiliation{Chonnam National University, Kwangju}
\affiliation{University of Cincinnati, Cincinnati, Ohio 45221}
\affiliation{University of Frankfurt, Frankfurt}
\affiliation{Gyeongsang National University, Chinju}
\affiliation{University of Hawaii, Honolulu, Hawaii 96822}
\affiliation{High Energy Accelerator Research Organization (KEK), Tsukuba}
\affiliation{Hiroshima Institute of Technology, Hiroshima}
\affiliation{Institute of High Energy Physics, Chinese Academy of Sciences, Beijing}
\affiliation{Institute of High Energy Physics, Vienna}
\affiliation{Institute for Theoretical and Experimental Physics, Moscow}
\affiliation{J. Stefan Institute, Ljubljana}
\affiliation{Kanagawa University, Yokohama}
\affiliation{Korea University, Seoul}
\affiliation{Kyungpook National University, Taegu}
\affiliation{Swiss Federal Institute of Technology of Lausanne, EPFL, Lausanne}
\affiliation{University of Ljubljana, Ljubljana}
\affiliation{University of Maribor, Maribor}
\affiliation{University of Melbourne, Victoria}
\affiliation{Nagoya University, Nagoya}
\affiliation{Nara Women's University, Nara}
\affiliation{National Central University, Chung-li}
\affiliation{National United University, Miao Li}
\affiliation{Department of Physics, National Taiwan University, Taipei}
\affiliation{H. Niewodniczanski Institute of Nuclear Physics, Krakow}
\affiliation{Nihon Dental College, Niigata}
\affiliation{Niigata University, Niigata}
\affiliation{Osaka City University, Osaka}
\affiliation{Osaka University, Osaka}
\affiliation{Panjab University, Chandigarh}
\affiliation{Peking University, Beijing}
\affiliation{University of Science and Technology of China, Hefei}
\affiliation{Seoul National University, Seoul}
\affiliation{Sungkyunkwan University, Suwon}
\affiliation{University of Sydney, Sydney NSW}
\affiliation{Tata Institute of Fundamental Research, Bombay}
\affiliation{Toho University, Funabashi}
\affiliation{Tohoku Gakuin University, Tagajo}
\affiliation{Tohoku University, Sendai}
\affiliation{Department of Physics, University of Tokyo, Tokyo}
\affiliation{Tokyo Institute of Technology, Tokyo}
\affiliation{Tokyo Metropolitan University, Tokyo}
\affiliation{Tokyo University of Agriculture and Technology, Tokyo}
\affiliation{University of Tsukuba, Tsukuba}
\affiliation{Virginia Polytechnic Institute and State University, Blacksburg, Virginia 24061}
\affiliation{Yonsei University, Seoul}
   \author{K.-F.~Chen}\affiliation{Department of Physics, National Taiwan University, Taipei} 
   \author{K.~Abe}\affiliation{High Energy Accelerator Research Organization (KEK), Tsukuba} 
   \author{K.~Abe}\affiliation{Tohoku Gakuin University, Tagajo} 
   \author{H.~Aihara}\affiliation{Department of Physics, University of Tokyo, Tokyo} 
   \author{Y.~Asano}\affiliation{University of Tsukuba, Tsukuba} 
   \author{V.~Aulchenko}\affiliation{Budker Institute of Nuclear Physics, Novosibirsk} 
   \author{T.~Aushev}\affiliation{Institute for Theoretical and Experimental Physics, Moscow} 
   \author{S.~Bahinipati}\affiliation{University of Cincinnati, Cincinnati, Ohio 45221} 
   \author{A.~M.~Bakich}\affiliation{University of Sydney, Sydney NSW} 
   \author{I.~Bedny}\affiliation{Budker Institute of Nuclear Physics, Novosibirsk} 
% ...
\collaboration{The Belle Collaboration}
       


\begin{abstract}
We present measurements of decay amplitudes and 
triple-product correlations in $B \to \phi K^*$ decays based on  
253~fb$^{-1}$ of data recorded at the $\Upsilon(4S)$ resonance with
the Belle detector at the KEKB $e^{+} e^{-}$ storage ring.
The decay amplitudes for the three different helicity states  
are determined from the angular distributions of final state particles. 
The longitudinal polarization amplitudes are 
found to be $0.45 \pm 0.05 \pm 0.02$ for $B^0 \to \phi K^{*0}$
and $0.52 \pm 0.08 \pm 0.03$ for $B^+ \to \phi K^{*+}$ decays.
$CP$- and $T$-odd $CP$-violating triple-product
asymmetries are measured to be consistent
with zero.
\end{abstract}
\pacs{13.25.Hw, 11.30.Er}

\maketitle

\tighten

{\renewcommand{\thefootnote}{\fnsymbol{footnote}}}
\setcounter{footnote}{0}







   
  The vector-vector $B\to\phi K^*$ decay processes
  provide clear insights into the underlying $b\to s$
  transition by virtue of their clear experimental
  signatures and relatively unambiguous theoretical
  interpretation, 	      
 especially of the many angular correlations
 that can be formed among the final-state particles.
 The decays are described by second order penguin diagrams, the first 
 order $b\to s$ transition being forbidden in the Standard Model (SM). 
 The angular information  allows the $CP$-even and
$CP$-odd states that comprise the $B^0 \to \phi K^{*0}$ decay to be distinguished.
 Our previous measurement \cite{ref:phik_prl} and a recent report by {\sc BaBar} \cite{ref:babar_ex0408017} 
 both suggest that the longitudinal polarization component differs from predictions based on the 
 factorization assumption.
 
 In this letter we report on a further study
 of this anomaly that is based on a larger
 data sample and uses observables that are
 expected to be sensitive to the effects of
 new physics.
 We present the first full three-dimensional angular analysis 
 for $B^+ \to \phi K^{*+}$  
 and an extended study for $B^0\to \phi K^{*0}$.
The decay modes $\phi \to K^+ K^-$, $K^{*0} \to K^+\pi^-$, $K^{*+} \to K^0_S\pi^+$, and 
$K^{*+} \to K^+\pi^0$ are considered.
Charge conjugate modes are implied everywhere unless otherwise specified.
We report measurements of direct $CP$ asymmetries, triple-product correlations and related $T$-odd $CP$-violating asymmetries \cite{ref:TP-datta},
and other observables that are sensitive to New Physics (NP) \cite{ref:NP-bounds}.


This analysis 
 uses a data sample that contains
 $275\times 10^6~B\overline{B}$ pairs collected
on the $\Upsilon(4S)$ resonance
by the Belle detector \cite{ref:Belle}
at the KEKB  $e^{+}e^{-}$ collider \cite{ref:KEKB}.
The Belle detector is a general purpose magnetic 
spectrometer equipped with a 1.5~T superconducting solenoid magnet. 
Charged tracks are reconstructed in a central drift chamber (CDC)
and a silicon vertex detector (SVD). 
Photons and electrons are identified using a CsI(Tl) electromagnetic calorimeter (ECL) 
located inside the magnet coil. 
Charged particles are identified using specific ionization ($dE/dx$)
measurements in the CDC as well as information from aerogel Cherenkov 
counters (ACC) and time of flight counters (TOF). 


Event reconstruction is performed as described in Ref.~\cite{ref:phik_prl}.
Candidate $B$ mesons are reconstructed from $\phi$ and $K^*$ candidates and are 
identified by the energy difference 
$\Delta E = E_B^{\rm cms} - E_{\rm beam}^{\rm cms}$, 
the beam constrained mass $M_{\rm bc} = \sqrt{(E^{\rm cms}_{\rm beam})^2 - (p_B^{\rm cms})^2}$,
and $K^+K^-$ invariant mass ($M_{K^+K^-}$), where
$E_{\rm beam}^{\rm cms}$ is the beam energy in the center-of-mass system (cms),
and $E_B^{\rm cms}$ and $p_B^{\rm cms}$ are the cms energy and momentum 
of the reconstructed $B$ candidate.
The $B$-meson signal region is defined as $M_{\rm bc} >$ 5.27 GeV/$c^2$, 
$|\Delta E| <$ 45 MeV, and $|M_{K^+K^-}-M_\phi| < 10$ MeV/$c^2$. 
The invariant mass of the $K^{*} \to K\pi$ candidate is required to be less than 
70 MeV/$c^2$ from the nominal $K^{*}$ mass.
The signal region is enlarged to $-100$ MeV $<\Delta E<$ $80$ MeV for 
$B^+ \to \phi K^{*+} (K^{*+} \to K^{+} \pi^{0})$ 
because of the effects of shower leakage on the $\Delta E$ resolution.
An additional requirement $\cos\theta_{K^*}<0.8$
is applied to reduce low momentum $\pi^0$ background,
where $\theta_{K^*}$ is the angle between the direction opposite
to the $B$ and the daughter kaon in the rest frame of $K^*$.
These requirements do not effect our results based on a MC study.
In the signal region, about 1\% of the events have multiple candidates.
The candidate with the smallest $\chi^2$ value from $B$ vertex finding
and the best $\pi^0$ mass in the $K^{*+}\to K^+\pi^0$ decay is used.


The dominant background is  
$e^+e^- \to q\overline{q}$ ($q=u,d,c,s$) continuum production.
Several variables including $S_{\perp}$ \cite{ref:sper}, 
the thrust angle, and the modified Fox-Wolfram
moments defined in Ref.~\cite{ref:continuum_suppression}
are used to exploit the differences between 
the event shapes for continuum $q\overline{q}$ production
(jet-like) and for $B$ decay (spherical) in the cms frame of 
the $\Upsilon(4S)$.
These variables are combined into a single likelihood ratio 
${\cal R}_s = {\cal L}_s/({\cal L}_s + {\cal L}_{q\overline{q}})$, where 
${\cal L}_s$ (${\cal L}_{q\overline{q}}$) denotes the signal (continuum)
likelihood. The selection requirements on ${\cal R}_s$ are determined 
by maximizing the value of $N_{s}/\sqrt{N_{s} + N_{b}}$
in each $B$-flavor-tagging quality region \cite{ref:tagging}, where
$N_{s}$ ($N_{b}$) represents the expected number of signal (background) events
in the signal region.


Backgrounds from other $B$ decay modes such as $B \to K^+K^-K^{*}$, 
$B \to f_0(980) K^{*}(f_0\to K^+K^-)$, $B \to \phi K\pi$, $B \to K^+K^-K\pi$, and 
cross-feed between the $\phi K^{*}$ and $\phi K$ decay channels are studied. 
The contributions from $B \to K^+K^-K^{*}$ and $B \to f_0 K^{*}$ 
are estimated from a fit to $\Delta E$, $M_{\rm bc}$, and 
$M_{K^+K^-}$ distributions.
The $M_{K^+K^-}$ distribution for $B \to K^+K^-K^{*}$ is determined from Monte Carlo (MC) simulations.
The $f_0(980)$ line shape is obtained from MC, where an $S$-wave Breit-Wigner with a
61 MeV/$c^2$ intrinsic width \cite{ref:garmash} is assumed. 
The uncertainty in the $f_0(980)$ width (40--100 MeV/$c^2$) \cite{ref:PDG} is taken
as a source of systematic error.
The contributions from $B \to K^+K^-K^{*}$ ($B \to f_0 K^{*}$) are estimated 
together with the $\phi K^*$ signal and are found to be
1~to 7\% (1~to 3\%) \cite{ref:range} of the signal yield.
The background from $B \to \phi K\pi$ decays is evaluated with fits to the $K\pi$ invariant mass
and is found to be about 1\%.
The contamination from four-body $B \to K^+K^-K\pi$ decays is checked by performing 
fits to the events in the $\phi \to K^+K^-$ and $K^* \to K\pi$ mass sidebands and is found to be less than 1\%. 
To remove the contamination from $\phi K$ decays, these decays are explicitly
reconstructed and rejected.


The signal yields ($N_{s}$) are extracted by 
extended unbinned maximum-likelihood fits 
performed simultaneously to the $\Delta E$, $M_{\rm bc}$ and $M_{K^+K^-}$ distributions.
Reconstructed $B$ candidates with $|\Delta E| < 0.25$ GeV, $M_{\rm bc} > 5.2$ GeV/$c^2$, and 
$M_{K^+K^-} < 1.07$ GeV/$c^2$ are included in the fits. 
The signal probability density
functions (PDFs) are a single Gaussian in $M_{\rm bc}$, 
a core Gaussian plus a Bifurcated Gaussian (Gaussian with different widths 
on either side of the mean) as the tail in the $\Delta E$ distribution, and 
a Breit-Wigner shape in $M_{K^+K^-}$. 
The means and widths of  $\Delta E$ and $M_{\rm bc}$ are 
verified using $B \to J/\psi K^*$ decays.
The mean and width of the $\phi$ mass peak are determined using an inclusive 
$\phi \to K^+K^-$ data sample.

\begin{figure}[!htb]
\centerline{
\hskip -0.4cm
\resizebox*{3.5in}{2.0in}{\includegraphics{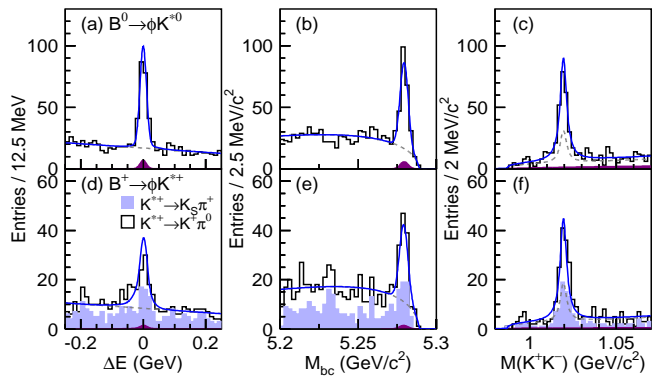}}}
\caption{ Distributions of the $\Delta E$, $M_{\rm bc}$, and  $M_{K^+K^-}$ for
$B^0\to \phi K^{*0}$ ((a),(b) and (c)), and for
$B^+\to \phi K^{*+}$ ((d),(e) and (f)), with other variables in signal
region. 
Solid curves show the fit results.
The continuum background components are shown by the dashed curves. 
The dark shaded areas represent the contributions from $B \to K^+K^-K^{*}$ and
$B \to f_0 K^{*}$ decays.}
\label{fig:demb-projections}
\end{figure}

The PDF shapes for the continuum events are parameterized by an ARGUS function \cite{ref:argus} in
$M_{\rm bc}$, a linear function in $\Delta E$, and a sum of a threshold function and a Breit-Wigner 
function in $M_{K^+K^-}$.
The parameters of the functions are 
determined by a fit to the events in the sideband.  
The signal and background yields are allowed to float in the fit while other PDF 
parameters are fixed.
The direct $CP$ asymmetries,
$A_{CP}=\frac{N(\overline{B} \to \overline{f})-N(B \to f)}
  {N(\overline{B} \to \overline{f})+N(B \to f)}$,
are also studied.
The measured signal yields and direct $CP$ asymmetries are summarized in Table~\ref{tab:yields}.
The distributions of $\Delta E$, $M_{\rm bc}$, and $M_{K^+K^-}$ are shown in Fig.~\ref{fig:demb-projections}.


\begin{table}[!htb]
\caption{Number of events observed in the signal region ($N_{\rm ev}$),   
signal yields ($N_{s}$) and the direct $CP$ asymmetries (${A}_{CP}$) obtained in the fits,
with statistical and systematic uncertainties.}
\label{tab:yields}	 
\begin{tabular}{l|ccc}
\hline 
\hline
Mode  			    & $N_{\rm ev}$ & $N_{s}$ & ${A}_{CP}$ \\
\hline
$\phi K^{*0}$         	    & 309  & $173 \pm 16$         & $0.02 \pm 0.09 \pm 0.02$ \\
\hline
$\phi K^{*+}$               & 173  & $85^{+12}_{-11}$ 	  & $-0.02 \pm 0.14 \pm 0.03$ \\
~~$K^{*+}(K^{0}_{S}\pi^{+})$  & 76   & $37.9^{+7.7}_{-7.0}$ & $-0.14 \pm 0.21 \pm 0.04$ \\
~~$K^{*+}(K^{+}\pi^{0})$      & 97   & $47.3^{+9.1}_{-8.1}$ & $ 0.09 \pm 0.19 \pm 0.04$ \\
\hline 
\hline
\end{tabular}
\end{table}

The decay angles of a $B$-meson decaying to two vector mesons $\phi$ and $K^{*}$ are defined
in the transversity basis \cite{ref:transversity}. 
The $x$-$y$ plane is defined to be the decay plane of $K^{*}$ and the $x$ axis is in
the direction of the $\phi$-meson.
The $y$ axis is perpendicular to the $x$ axis in the decay plane and is on the same side as the kaon from the $K^{*}$ decay.
The $z$ axis is perpendicular to the $x$-$y$ plane according to the right-hand rule,
$\theta_{\rm tr}$ ($\phi_{\rm tr}$) is the polar (azimuthal) angle 
with respect to the $z$-axis of the $K^+$ from $\phi$ decay 
in the $\phi$ rest frame, and $\theta_{K^*}$ is defined earlier.

The distribution of the angles, $\theta_{K^*}$, $\theta_{\rm tr}$, and
$\phi_{\rm tr}$ is given by \cite{ref:polarization_hepex}
\begin{eqnarray}
\label{equ:angularpdf}
\nonumber
&&{d^3 R_{\phi K^{*}} (\phi_{\rm tr}, \cos\theta_{\rm tr}, \cos\theta_{K^*}) \over d\phi_{\rm tr} d\cos{\theta_{\rm tr}} d\cos{\theta_{K^*}}} 
= {9\over 32\pi} [ \\
\nonumber
&&~~~~~~~~ \phantom{+}~ |A_\perp|^2 2 \cos^2{\theta_{\rm tr}} \sin^2{\theta_{K^*}} \\
\nonumber
&&~~~~~~~~ +|A_\parallel|^2 2 \sin^2{\theta_{\rm tr}} \sin^2{\phi_{\rm tr}} \sin^2{\theta_{K^*}} \\
\nonumber
&&~~~~~~~~ + |A_0|^2 4 \sin^2\theta_{\rm tr} \cos^2\phi_{\rm tr} \cos^2\theta_{K^*} \\
\nonumber
&&~~~~~~~~ +\sqrt{2} {\rm Re}(A^*_\parallel A_0) \sin^2\theta_{\rm tr}\sin 2 \phi_{\rm tr} \sin 2\theta_{K^*} \\
\nonumber
&&~~~~~~~~ -\eta\sqrt{2} {\rm Im}(A_0^* A_\perp) \sin 2\theta_{\rm tr} \cos\phi_{\rm tr} \sin 2\theta_{K^*} \\
&&~~~~~~~~ -2\eta{\rm Im}(A_\parallel^*A_\perp) \sin 2 \theta_{\rm tr} \sin \phi_{\rm tr} \sin^2 \theta_{K^*} ]~,
\end{eqnarray}
where $A_0$, $A_\parallel$, and $A_\perp$ are the complex amplitudes of the
three helicity states in the transversity basis with the normalization condition
$|A_0|^2 + |A_\parallel|^2 + |A_\perp|^2 = 1$, and $\eta = +1$ ($-1$) corresponds to
$B$ ($\overline{B}$) mesons and is determined from the charge of the kaon or pion in the $K^{*}$ decay.
The longitudinal polarization component is denoted by $A_0$;
$A_\perp$ ($A_\parallel$) is the transverse polarization
along the $z$-axis ($y$-axis).
The value of $|A_\perp|^2$ ($|A_0|^2 + |A_\parallel|^2$)
is the $CP$-odd ($CP$-even)
fraction in the decay $B \to \phi K^{*0}$ \cite{ref:polarization_hepex}. 
The presence of final state interactions (FSI) results in phases that differ from either $0$ or $\pm\pi$.

\begin{figure}[!htb]
\centerline{
\resizebox*{3.2in}{2.0in}{
\includegraphics{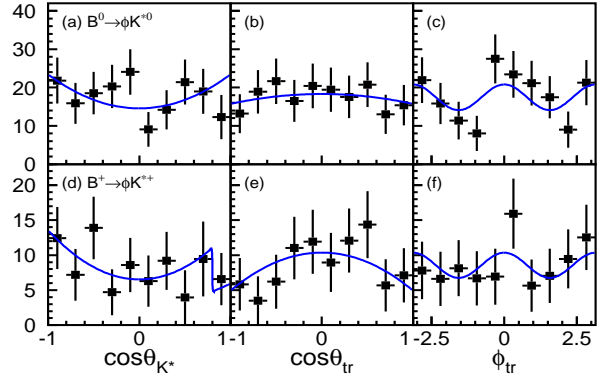}}
}   
\caption{ 
Projected distributions of the three transversity angles for
$B^0\to \phi K^{*0}$ ((a),(b) and (c)), and for
$B^+\to \phi K^{*+}$ ((d),(e) and (f)).
Solid lines show the fit results.
The points with error bars show the 
efficiency corrected data after background subtraction.
The two $K^{*+}$ decay modes are combined in (d),(e) and (f). 
The discontinuity in (d) is due to the requirement of $\cos\theta_{K^*} < 0.8$
in $B^+ \to\phi K^{*+}(K^{*+}\to K^+\pi^0)$.}
\label{fig:angular-proj}
\end{figure}

The complex amplitudes are determined by performing
an unbinned maximum likelihood fit  
to the $B \to \phi K^{*}$ candidates in the signal region.
The combined likelihood is given by
\begin{equation}
\mathcal{L} =
\prod_i^{N_{\rm ev}} \epsilon(\phi_{\rm tr}, \cos\theta_{\rm tr}, \cos\theta_{K^*}) 
\sum_j f_{j} R_{j}(\phi_{\rm tr}, \cos\theta_{\rm tr}, \cos\theta_{K^*})~,
\end{equation}
where $j$ denotes the contributions from 
$\phi K^*$, $q\overline{q}$, $K^+K^-K^*$ and $f_0K^*$; 
$R_{j}$ is the angular distribution function (ADF).
The ADF $R_{q\overline{q}}$ is determined from sideband data, and 
$R_{K^+K^-K^{*}}$ from events with 1.04 GeV/$c^2$ $<$ $M_{K^+K^-}$ $<$ 1.075 GeV/$c^2$;
$R_{f_0K^{*}}$ is obtained from $B \to f_0 K^*$ MC events.
The detection efficiency ($\epsilon$) is determined using MC simulations
assuming a phase space decay.
The fractions $f_j$ are parameterized as 
a function of $\Delta E$, $M_{\rm bc}$ and $M_{K^+K^-}$.
The value of $\arg(A_0)$ is set to zero and 
$|A_\parallel|^2$ is calculated from the normalization condition.
The four parameters ($|A_0|^2$, $|A_\perp|^2$, $\arg(A_\parallel)$, and $\arg(A_\perp)$)
are determined from the fit.
There is a two-fold ambiguity in the solutions for 
the phases; the chosen set of solutions is the one suggested in Ref.~\cite{ref:2solutions}.
Figure~\ref{fig:angular-proj} shows the angular distributions with projections
of the fit superimposed.
The obtained amplitudes are summarized in Table~\ref{tab:amplitudes}.

The systematic uncertainties on the amplitudes are dominated by 
the efficiency modeling (4--5\%), continuum background (3--4\%),
slow pion efficiency (2--3\%), and $K^+K^-K^*$ ADF (1--2\%). The remaining possible
systematic errors, such as the angular resolution, signal yields, 
background from higher $K^*$ states, and width of the $f_0$, are estimated to be 
less than 1\%.  


\begin{table}[!htb]
\caption{The decay amplitudes obtained for $B^0\to \phi K^{*0}$
and $B^+\to \phi K^{*+}$.
The first uncertainties are statistical and the second are systematic.}
\label{tab:amplitudes}
\begin{center}
\begin{tabular}{c|cc}
\hline 
\hline 
Mode & $\phi K^{*0}$ & $\phi K^{*+}$ \\
\hline
$|A_0|^2$	         & $0.45 \pm 0.05 \pm 0.02$  &~~ $0.52 \pm 0.08 \pm 0.03$  \\
$|A_\perp|^2$ 	         & $0.30 \pm 0.06 \pm 0.02$  &~~ $0.19 \pm 0.08 \pm 0.02$  \\
$\arg(A_\parallel)$ (rad) & $2.39 \pm 0.24 \pm 0.04$  &~~ $2.10 \pm 0.28 \pm 0.04$ \\
$\arg(A_\perp)$ (rad)     & $2.51 \pm 0.23 \pm 0.04$  &~~ $2.31 \pm 0.30 \pm 0.07$  \\
\hline
\hline
\end{tabular}
\end{center}
\end{table}


The triple-product for a $B$ meson decay to two vector mesons takes the form 
$\overrightarrow{q} \cdot (\overrightarrow{\epsilon_1} \times \overrightarrow{\epsilon_2})$,
where $\overrightarrow{q}$ is the momentum of one of the vector mesons, and
$\overrightarrow{\epsilon_1}$ and $\overrightarrow{\epsilon_2}$ are 
the polarizations of the two vector mesons. 
The following two $T$-odd \cite{ref:TP-datta,ref:at12} quantities
\begin{equation}
A_T^{0} = {{\rm Im}(A_\perp A^*_0) }~,
~~~~
A_T^{\parallel} = {{\rm Im}(A_\perp A^*_\parallel) }~,
\end{equation}
provide information on the asymmetry of the triple products.
The SM predicts very small values for $A_T^{0}$ and $A_T^{\parallel}$.
The comparison of these triple product asymmetries ($A_T^{0}$ and $A_T^{\parallel}$)
with the corresponding quantities for the $CP$-conjugate decays 
($\overline{A}_T^{0}$ and $\overline{A}_T^{\parallel}$) provides an 
observable sensitive to $T$-odd $CP$-violation. 

Additional variables that can be accessed by angular analyses 
are suggested in Ref.~\cite{ref:NP-bounds} and are given by
\begin{eqnarray}
\nonumber
&& \hskip -0.4cm
\Lambda_{\perp i} = -{\rm Im} ( A_\perp A_i^* - \overline{A}_\perp \overline{A}_i^* ),~  
\Lambda_{\parallel 0} = {\rm Re} ( A_\parallel A_0^* + \overline{A}_\parallel \overline{A}_0^* ),  
\\
\nonumber
&& \hskip -0.4cm
\Sigma_{\perp i} = -{\rm Im} ( A_\perp A_i^* + \overline{A}_\perp \overline{A}_i^* ),~  
\Sigma_{\parallel 0} = {\rm Re} ( A_\parallel A_0^* - \overline{A}_\parallel \overline{A}_0^* ),
\\
&& \hskip -0.4cm
\Lambda_{\lambda\lambda} = {1\over 2} ( |A_\lambda|^2 + |\overline{A}_\lambda|^2  ),~
\Sigma_{\lambda\lambda} = {1\over 2} ( |A_\lambda|^2 - |\overline{A}_\lambda|^2  ),
\end{eqnarray}
where the subscript $\lambda$ is either 0, $\parallel$, or $\perp$,
and $i$ is 0 or $\parallel$. 
The variables $\Lambda_{\perp 0}$ and $\Lambda_{\perp\parallel}$ are sensitive to
$T$-odd $CP$-violating new physics.
The following equations should hold 
in the absence of NP:
\begin{eqnarray}
\Sigma_{\lambda\lambda} = 0~, &
\Sigma_{\parallel 0} = 0~, &
\Lambda_{\perp i} = 0~.
\end{eqnarray}

By separating $B^0$ and $\overline{B}{}^0$ samples and rearranging fitting parameters 
in the unbinned maximum likelihood fit, we measured the decay amplitudes for the $B^0$ and $\overline{B}{}^0$,
the triple-product correlations, and the other NP-sensitive observables as given in
Table~\ref{tab:amplitudes-sep} and \ref{tab:NP-observables}.  
The $T$-odd $CP$-violating variables 
$\Lambda_{\perp 0}$ and $\Lambda_{\perp\parallel}$
are measured to be
$0.16^{+0.16}_{-0.14}\pm0.03$ and $0.01\pm0.10\pm0.02$, respectively, 
consistent with the SM predictions.


\begin{table}[!htb]
\caption{The measured decay amplitudes and triple-product correlations
in the $B^0$ and $\overline{B}{}^0$ samples.}
\label{tab:amplitudes-sep}
\begin{center}
\begin{tabular}{c|cc}
\hline 
\hline 
Mode & $B^0$  & $\overline{B}{}^0$  \\
\hline
$|A_0|^2$	         & $0.39 \pm 0.08 \pm 0.03$         & $0.51 \pm 0.07 \pm 0.02$ \\
$|A_\perp|^2$ 	         & $0.37 \pm 0.09 \pm 0.02$         & $0.25 \pm 0.07 \pm 0.01$ \\
$\arg(A_\parallel)$ (rad) & $2.72_{-0.38}^{+0.46} \pm 0.14$  & $2.08 \pm 0.31 \pm 0.04$ \\
$\arg(A_\perp)$ (rad) 	 & $2.81 \pm 0.36 \pm 0.11$         & $2.22 \pm 0.35 \pm 0.05$ \\
\hline
$A_T^{0}$         & $0.13^{+0.11}_{-0.14}\pm0.04$  & $0.28\pm0.08\pm0.01$ \\
$A_T^{\parallel}$ & $0.03\pm0.08\pm0.01$  & $0.03\pm0.06\pm0.01$ \\
\hline
\hline
\end{tabular}
\end{center}
\end{table}


\begin{table}[!htb]
\caption{$\Lambda$ and $\Sigma$ values obtained from the
decay amplitudes measured for $B^0$ and $\overline{B}{}^0$ separately.}
\label{tab:NP-observables}
\begin{center}
\begin{tabular}{lcr|lcr}
\hline 
\hline 
$\Lambda_{00}$  	       &=& $ 0.45\pm0.05\pm0.02$          & $\Sigma_{00}$		  &=& $-0.06\pm0.05\pm0.01$ \\
$\Lambda_{\parallel\parallel}$ &=& $ 0.24\pm0.06\pm0.02$          & $\Sigma_{\parallel\parallel}$ &=& $-0.01\pm0.06\pm0.01$ \\
$\Lambda_{\perp\perp}$         &=& $ 0.31\pm0.06\pm0.01$          & $\Sigma_{\perp\perp}$	  &=& $ 0.06\pm0.05\pm0.01$ \\
$\Sigma_{\perp 0}$	       &=& $-0.41^{+0.16}_{-0.14}\pm0.04$ & $\Lambda_{\perp 0}$	  &=& $ 0.16^{+0.16}_{-0.14}\pm0.03$ \\
$\Sigma_{\perp\parallel}$      &=& $-0.06\pm0.10\pm0.01$ 	  & $\Lambda_{\perp\parallel}$    &=& $ 0.01\pm0.10\pm0.02$	     \\
$\Lambda_{\parallel 0}$        &=& $-0.45\pm0.11\pm0.01$          & $\Sigma_{\parallel 0}$	  &=& $-0.11\pm0.11\pm0.02$ \\
\hline
\hline
\end{tabular}
\end{center}
\end{table}


 In summary, improved measurements of the decay 
 amplitudes for $B\to\phi K^*$, based on fits to
 angular distributions in the transversity basis,
 are presented.
 The results are consistent with our previous measurements~\cite{ref:phik_prl} with
 improved precision. 
The measured value of $|A_\perp|^2$ 
shows that $CP$-odd ($|A_\perp|^2$) and $CP$-even ($|A_0|^2 + |A_\parallel|^2$) 
components are present in $\phi K^*$ decays in a ratio of about 1:2.
Phases of both $A_\perp$ and $A_\parallel$ differ from zero or $-\pi$ 
by 4.3 standard deviations ($\sigma$), which provides evidence for
the presence of final state interactions.
The measured direct $CP$ asymmetries in these modes are consistent with zero;
the corresponding 90\% confidence level limits are
$-0.14 < A_{CP}(\phi K^{*0}(K^+\pi^-)) < 0.17$, 
and $-0.25 < A_{CP}(\phi K^{*+}) < 0.22$.
  Measurements of the $T$-odd $CP$-violation sensitive 
  differences between triple product asymmetries,
  $A^0_T - \overline{A}^0_T$  and $A^{\parallel}_T - \overline{A}^{\parallel}_T$, indicate 
  no significant deviations from zero, consistent with {\sc BaBar} measurements~\cite{ref:babar_ex0408017}. 
  Our data shows no significant deviations from the expectations: 
   $\Sigma_{\lambda\lambda} = 0$, $\Sigma_{\parallel 0} = 0$, and
   $\Lambda_{\perp i} = 0$, indicating no
  evidence for new physics. 


We thank the KEKB group for the excellent operation of the
accelerator, the KEK cryogenics group for the efficient
operation of the solenoid, and the KEK computer group and
the NII for valuable computing and Super-SINET network
support.  We acknowledge support from MEXT and JSPS (Japan);
ARC and DEST (Australia); NSFC (contract No.~10175071,
China); DST (India); the BK21 program of MOEHRD and the CHEP
SRC program of KOSEF (Korea); KBN (contract No.~2P03B 01324,
Poland); MIST (Russia); MHEST (Slovenia);  SNSF (Switzerland); NSC and MOE
(Taiwan); and DOE (USA).



\begin{thebibliography}{99}
\bibitem{ref:range}
{The range corresponds to the range of values for different
sub-modes. This convention is used throughout this letter.}
\bibitem{ref:at12}
{We take the definitions of $A_T^0 = A_T^{(1)}$ and 
$A_T^\parallel = A_T^{(2)}$, but $\overline{A}_T^0 = -\overline{A}_T^{(1)}$ and 
$\overline{A}_T^\parallel = -\overline{A}_T^{(2)}$. The variables $A_T^{(1)}$ and $A_T^{(2)}$
are defined in Ref.~\cite{ref:TP-datta}.}
\end{thebibliography}
\end{document}